\documentclass[
aps,
12pt,
final,
notitlepage,
oneside,
nobibnotes,
nofootinbib,
superscriptaddress,
showpacs]
{revtex4}
\usepackage{graphicx}
\usepackage{amsmath}
\usepackage{amsfonts}
\usepackage{amssymb}
\usepackage{bbm}
\usepackage{young}
\newcommand{\PL}{\protect{Pl{\"u}cker }}
\newcommand{\PLL}{\protect{Pl{\"u}cker}}
\begin{document}
\title{On a Microscopic Representation of Space-Time IV\protect{\footnote{The 
author thanks the Alexander von Humboldt-Foundation (Bonn, Germany) for 
financial support.}}}
\author{\firstname{Rolf}~\surname{Dahm}}
\affiliation{Permanent address: beratung f{\"{u}}r Informationssysteme und Systemintegration,
G{\"{a}}rtnergasse 1, D-55116 Mainz, Germany}

\begin{abstract}
We summarize some previous work on SU(4) describing hadron representations 
and transformations as well as its non-compact 'counterpart' SU$*$(4) being
the complex embedding of SL(2,$\mathbb{H}$). So after having related the 
16-dim Dirac algebra to SU$*$(4), on the one hand we have access to real,
complex and quaternionic Lie group chains and their respective algebras,
on the other hand it is of course possible to relate physical descriptions
to the respective representations. With emphasis on the common maximal 
compact subgroup USp(4), we are led to projective geometry of real 3-space
and various transfer principles which we use to extend previous work on
the rank 3-algebras above. On real spaces, such considerations are governed
by the groups SO($n$,$m$) with $n+m=6$. The central thread, however, focuses
here on line and Complex geometry which finds its well-known counterparts in 
descriptions of electromagnetism and special relativity as well as - using
transfer principles - in Dirac, gauge and quantum field theory. We discuss
a simple picture where Complexe of second grade play the major and dominant
r\^{o}le to unify (real) projective geometry, complex representation theory
and line/Complex representations in order to proceed to dynamics.
\end{abstract}


\maketitle

\section{Introduction}
In \cite{dahm:2008} and \cite{dahm:MRST1} we have given arguments
why and how to treat the Dirac algebra and various of its aspects
in terms of groups on real, complex and quaternionic representation
spaces. There, we have identified spin and isospin degrees of freedom
within the compact group SU(4) ($A_{3}$) and it's representations,
and we have related the non-compact group SU$*$(4), emerging from
the complex embedding of quaternions, to the Dirac algebra via the
identifications $\gamma^{0}=i\mathcal{Q}_{30}$, 
$\gamma^{j}=-i\mathcal{Q}_{2j}$ and $\gamma_{5}=i\mathcal{Q}_{10}$,
$Q_{\alpha\beta}$ denoting group elements, generators of the Lie
algebra su$*$(4) or elements from the five-dimensional Riemannian
space SU$*$(4)/USp(4) or from the vector space $p$, su$*$(4)$= h+p$,
$h=$usp(4), according to the respective context(s). In \cite{dahm:MRST3},
using the Lie algebras of \cite{dahm:MRST1} as an intermediary step 
towards projective and line geometry, we have extended the above 
approach, originally given in terms of (transformation) groups and 
their (Lie) algebras, by relating those algebras to a geometrical 
counterpart in terms of projective and especially line and 
Complex\footnote{We use the terminology proposed in \protect{\cite{dahm:MRST3}} 
which has been introduced by \PL, and we denote by 'Complex' (with 
capital 'C') a line complex in order to avoid confusion with complex
numbers.} geometry.\\
Here, we want to summarize in section \ref{sec:road} briefly some 
of the major arguments developed so far and arrange them in a logic 
sequence in order to serve as a twofold basis for upcoming work 
and discussions:
On one hand, we can reduce the group theoretical, algebraical and
differential geometrical models which have been discussed over the
years to their original content and concept which is based on nothing
but Klein's Erlanger program. This means, a geometry is determined 
by the identification of states and the transformation groups, 
however, there exist various (geometrical) representations which
can be (inter-)related using transfer principles. Practically, a
single algebraic rep or equation can have {\it different} content
and meaning depending e.g.~on the interpretation of the coordinates
involved.
A lot of such representations and related transfer principles have
been discussed in literature, we have mentioned some examples in 
\cite{dahm:MRST3} (see also references). Within this setting, we
subsume complex representation theory and quantum field theory as
subsets obtained by using appropriate complexifications of line 
geometry and projective (point and line) coordinates.
On the other hand, it is known for more than a century (see 
e.g.~\cite{klein:1928} and references therein) how to derive euclidean,
elliptic and hyperbolic geometry (and as such their transformation
groups like a 'Lorentz' group, a 'Poincar\'{e}' group or a 'Galilei'
group or homogeneous transformations on appropriate representation
spaces as well) from projective geometry\footnote{Here, we do not
want to discuss more details or geometrical limites like contractions 
and expansions to relate homogeneous and affine transformations with 
respect to the 'Poincar\'{e}' group but we simply refer to \protect{\cite{gilmore:1974}}, 
chapter 10 and references therein, as an overview and providing some
examples and related algebra.}.

That is why in section \ref{sec:picture} we use a {\it very} simple
picture which helps to illustrate and understand the coincidence of 
various models and representations used so far. It is able to illustrate
the concepts and derive physical aspects of dynamical systems for 
later use.

\section{On the Road to Projective and Line Geometry}
\label{sec:road}
Thinking of simple dynamical systems, the driving terms of almost 
all dynamical models are typically twofold: We choose a basic setup
or description of the system in terms of one or more (then interacting
or coupled) usually (irreducible) representations (later for short
'irreps') and assume a certain dynamical behaviour governed by certain
group or algebra transformations, thus intrinsically assuming a 
'well-behaved' nature of the dynamical system. The typical representation
theory is thus based on Klein's Erlanger program, especially when 
using derivatives and analyticity to formulate the dynamics and 
system behaviour in terms of point manifolds and differential 
geometry. In this picture, the initial and the final state may
both be characterized in terms of the chosen representation(s)
(later 'reps') thus implicitly assuming a valid global coordinatization
of the process\footnote{Or at least some knowledge on how to relate
the coordinate systems used throughout the process/description.}.
Dynamics is described by applying a transformation (or a series
of transformations completely described e.g.~by Lie algebra 
generators or group actions) to the initial state, and applying
the transformations is sufficient to 'reach' and characterize
the final state in well-behaved rep spaces. Hence, in such 
scenarios the usual way to investigate dynamics by applying 
(Lie) group or algebra theory, invariant theory and appropriate
algebraic concepts often is sufficient and helpful, especially
when working with compact groups. However, non-compact groups 
and non-existing {\it global} coordinate systems complicate 
this algebraically founded approach considerably, and it helps
to go back to the very foundations and use geometric pathways 
to reestablish physical reasoning and avoid some of the 'rep-only'
or 'rep-induced' problems when having chosen not the best or 
most exact rep. The major (and almost the only) physical driving
term is the terminology of an 'action' (or 'energy') -- its 
formulation and conservation (or non-conservation) of energy 
-- which allows to apply certain formalisms. In standard text
books, this treatment is usually mapped to applying differential
geometry in terms of Hamilton and/or Lagrange formalism and 
appropriate differential equations. Group schemes and applications
considering energy as a conserved quantity usually apply 
representation theory of compact groups to physical problems
while identifying states with real or complex irreps -- thus
introducing 'necessary' involutions (like complex conjugates,
adjoints, hermiticity, \ldots) and the related metrical 
properties by hand. Please note, that most of such properties
from the viewpoint of projective geometry are {\it derived} 
features, i.e.~they are already contained in projective geometry
and they are derived typically by applying constraints on 
geometrical settings, objects, states and groups, e.g.~by 
certain geometrical limites or identifications of elements 
and objects.

Now we do not argue to abandon this approach but it is well-known
(although almost forgotten) that equivalent descriptions are possible,
mostly due to the fact that Lie theory is a special (polar) and 
subsidiary concept of line and/or projective geometry in that we
investigate tangents to a point manifold and their related dynamics,
or 'line elements' as unions of 'points' and lines\footnote{In euclidean
\protect{$\mathbb{R}^3$}, we may represent such efforts by choosing
3-dim point reps \protect{$\vec{x}=(x,y,z)$} whereas the line elements
are represented by 5-dim elements \protect{$(x,y,z,\mathrm{d}x:\mathrm{d}y:\mathrm{d}z)$},
i.e.~it is the {\it ratios} of the elements \protect{$\mathrm{d}x,\mathrm{d}y,\mathrm{d}z$}
which enter theory and describe the direction of the line intersecting
the point \protect{$\vec{x}$}; a concept which leads to Monge's and
Pfaff's equations (see \protect{\cite{dahm:MRST3}}).}. So instead
of being concerned of point manifolds only (in terms of projective 
geometry: investigating the orders of the curves), we can as well 
look for the tangents more general as lines and their behaviour (in 
projective geometry e.g.~by the classes of the curves or surfaces, 
by polar behaviour and duality or a possible projective generation
of those objects) or even use and relate both pictures as has been
suggested by \PL e.g.~in terms of his dimensional formulae 
\cite{plueckerNG:1868}. A simple example has been referenced in 
\cite{dahm:MRST3} where the representation (or requirement) of a 
quadratic Complex automatically yields light cones when expressed 
in terms of point coordinates (of the manifold). In other words, 
representing physical objects by quadratic Complexe (i.e.~Complexe
of second grade) yields objects like (second-order) cones and 'light 
cones' in a manifold or point picture {\it automatically}, as has 
been structurally required by \cite{ehlers:1972}. Please recall that 
typically light cones are introduced ad hoc by special relativity
in conjunction with Weyl's gauge philosophy, and invariances and 
'metrics' like the 'Lorentz metric' in special relativity require
an 'extension' of the mathematical standard definition of a metric
and a separate, non-standard treatment of negative metric values. 
This can be avoided using projective geometry and line (or better 
Complex) representations as well as Complex geometry. As such, 
Dirac's square root of the energy written in terms of momenta 
(for the free part) or in terms of line coordinates, respectively,
is a simple square of a line (or Complex) square, hence we find 
(linear 4-dim) transformation group reps acting on line reps 
\cite{dahm:MRST3}. This yields exactly our statement above relating
the Dirac matrices to linear reps of SU$*$(4) or SO(5,1), and 
we can use group and representation theory to perform the 
algebra and provide analytic reps. The 'potential' part may 
be formulated as well keeping in mind that we have to respect
normals\footnote{Differential geometry uses gradients and 
(potential) functions to represent normal behaviour within
the differential formalism by using the fact that a differential
operator diminishes the power of a point/manifold coordinate 
by 1. The same happens geometrically when introducing normal 
vectors by \protect{$n_{i}~\sim\epsilon_{ijk}x_{i}y_{k}$},
i.e.~a (sub-)determinant. Please note however, that {\it both}
reps above work {\it only} in {\it affine} or non-homogeneous 
euclidean coordinates! Hence the discussion of polar and
axial vectors and 'parity' discussions at a first glance 
are related to this euclidean/affine picture only! A more
sophisticated treatment can also investigate the Complex
of normals of point curves or appropriate surfaces.}, too, 
which can be included in line geometry using a simple 
euclidean/affine six-vector representation of a line which 
decomposes into a 3-dim polar vector and a 3-dim axial 
vector having different orders in terms of point coordinates 
(see e.g.~\cite{dahm:MRST3}, eq.~(I.B (2)), and references 
therein). This, however, is achieved having chosen a special 
geometrical setup by fixing geometrical objects which at the 
same time restricts (real) projective transformations to 
orthogonal ones, i.e.~we have introduced 'invariant' or 
'fixed' (geometrical) objects with respect to the respective 
(coordinate) transformations. So we feel free to choose 
(from our viewpoint) better suited reps, and we understand
point and manifold discussions subsequently as a subsidiary
concept\footnote{Axiomatically, a point is fixed by incidence
of two lines.} only. However, at any time it is possible 
to switch back to differential geometry and symmetric 
spaces \cite{dahm:2008}.

From above, we follow the reasoning and interpretation of 
tangent 'objects' in order to map them on lines or Complexe.
As an example, we use the global Riemannian space SU$*$(4)/USp(4)
(see \cite{dahm:2008}, sec. 2, eq.~(3)) with the rep
\begin{equation}
\label{eq:cosetrep}
X\,:=\,\exp V\,=\,\cosh\|x\|\,\mathbf{1}\,+\,\sinh\|x\|\frac{x_{a}}{\|x\|}\,
\mathcal{P}_{a}\,,
\end{equation}
where $\mathcal{P}_{a}\,,1\leq a\leq 5$, denote the five non-compact
generators of the reductive Lie algebra decomposition of su$*$(4)
when subtracting the ten (compact) usp(4) generators. Now with respect
to the interpretation of the coset coordinates in eq.~(\ref{eq:cosetrep})
we feel free to interpret the $\mathcal{P}_{a}$ as (generalized)
velocity operators causing the symmetry transformations and to 
relate them to (infinitesimal) line representations\footnote{Which,
however, have to be identified only after having complexified some generators 
and compared to an SO(3,3) \PL rep or the SO(6) Klein rep. In 
general, we can thus obtain the whole series \protect{SO($n$,$m$),
$n+m=6$}, as can be seen from the signature of the squares. However,
the respective interpretation of the coordinates has to be changed 
appropriately and carefully! Please note, that 'naive' complexification
of the coordinates usually leads to \protect{SU($n$,$m$), $n+m=4$},
but SO(5,1) is related to the quaternionic embedding SU$*$(4) and 
as such represents a special case.}. The symmetry (or 'invariance')
group has been shown to be SO(5,1) which reflects in hyperbolic 
functions and the rank 1, negatively curved 'space' (\cite{dahm:2008} 
and references).

From the viewpoint of dynamics in the underlying manifold, we
may use the Lie algebra to develop the point motion by 
transforming 'point' and tangent (or the 'point velocity') with
appropriate (Hamiltonian) symmetry constraints. On the other 
hand, we may take the viewpoint of Complexe (e.g.,~a (quadratic)
tangential Complex or a tetrahedral Complex), abandon the 
construction of manifolds by points, continuity and analyticity, 
and instead switch to the associated notion of classes and 
envelopping structures of such a manifold where polarity and 
dimensional formulae enter. There exist, as we'll see in section
III, also possibilities to construct or generate curves and
surfaces in projective geometry. So discussing projective 
geometry of $\mathbb{R}^{3}$, (self-dual or conjugated) lines,
Complexe and Complex geometry seem best suited as an underlying
framework. There, it is well-known that a representation in 
terms of point/manifold coordinates has to be expressed by
four (real) homogeneous coordinates whereas Complexe right from
the beginning have different, even higher-dimensional reps in 
terms of (antisymmetric) line coordinates, and the condition 
of a quadratic Complex comprises automatically a light cone 
rep in point coordinates. Introducing a 'time' coordinate 
in the coordinate rep of the underlying point manifold (thus 
side-by-side the concept of a vectorial 'velocity' and the
notion of dynamics in this local setup), we can identify $c$
(interpreted as 'infinity' in {\it velocity space}) in the 
light cone definition, and by defining inhomogeneous/affine
point coordinates we obtain ratios of velocities \cite{dahm:MRST3}
(for overall/equal local time) which is consistent with 
Gilmore's presentation \cite{gilmore:1974} of the typical 
coset constructions of representation spaces of transformation 
groups or symmetric spaces. The point coordinates themselves 
can be extracted only from this (euclidean) dynamical picture
by introducing a transformation parameter $t$ common to all
local coordinates $x_{i}~\sim v_{i}t$ where $v_{i}$ are the 
transformation parameters or (generalized) velocities\footnote{Or
using, more generally, projective (4-dim) transformations relating 
\protect{{\it different homogeneous}} coordinates \protect{$x_{\alpha}$}
while avoiding an overall '$t$'. Anyhow, all approaches using linear
'time' have to reflect and represent 'velocity' equivalence classes
from (special) relativity with respect to coordinates in projective
geometry appropriately, and linear 'time' (parameter) dependence
may be even replaced by more complicated dependencies.}.
So from the viewpoint of an overall picture, we may use different
$t$-values to enumerate different consecutive geometrical point 
sets of the dynamical system in the manifold. Hence we can label
and relate the individual system states time $t_{n}$ by time 
$t_{n+1}$ from an overall (static) and purely geometrical setup
while assuming appropriate transformations connecting the labeled
states. The velocities (respectively their ratios) for the same 
time value reflect the r\^{o}le of building equivalence classes 
like the homogeneous coordinates of the point picture.
This, on the other hand, is consistent with the group picture
and the rep construction scheme using Lie algebras where 
one-parameter transformation groups transform and develop 
the physical/dynamical system \cite{gilmore:1974}.

\section{A Simple Picture Departing from Point Geometry}
\label{sec:picture}
For us, before going too much into detailed analytical discussions,
a simple underlying picture departing from the usual point (or 
manifold) picture allows to attach and construct the various 
representations and models, if we start from nothing but two 
individual and distinguishable points $x_{1}$ and $y_{1}$ (see 
figure 1a).
\begin{figure}[h]
\includegraphics[scale=0.55]{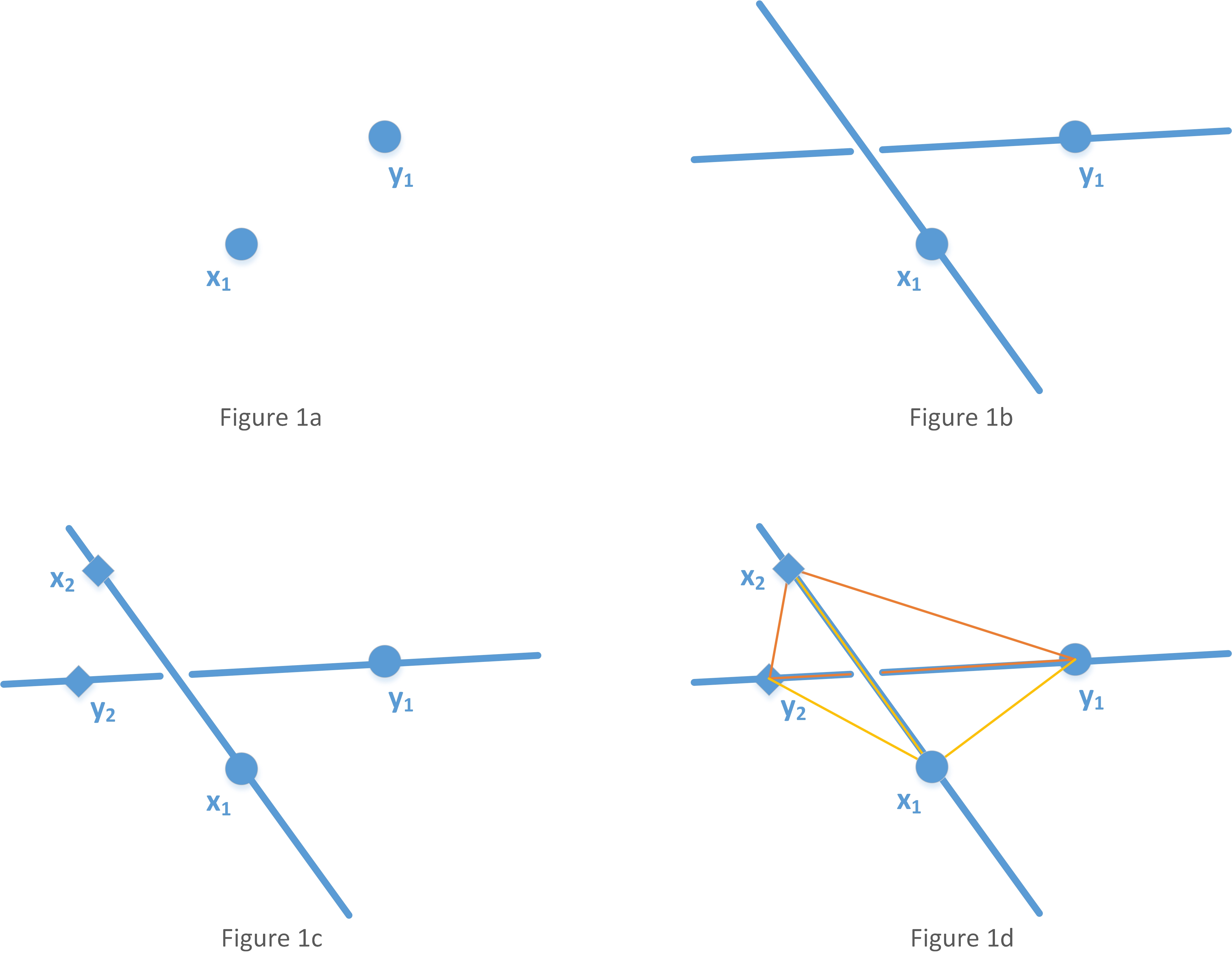}
\caption{Linear motion of points}
\label{fig:fig1}
\end{figure}

Taking the points $x_{1}$ and $y_{1}$ for given (represented by circles)
and transforming them individually by some (linear) action (figure 1b) 
to {\it different} points $x_{2}$ and $y_{2}$ (represented by diamonds),
we may represent each connection of the two points in the point sets 
$x_{1}\longrightarrow x_{2}$ and $y_{1}\longrightarrow y_{2}$,
generated by the (symmetry) transformation, by a line (figure 1c). Then
already this very basic assumption comprises a lot of information and
content in order to attach a wealth of formalism as we know from analysis
and differential geometry.

Here however, and that's why we have drawn figure 1d, we gain a fortune
in that the tetrahedron (which results when connecting the four points
by lines) yields a foundation of connecting three (analytic) approaches
we have used before only separately\footnote{We are aware of richer 
mechanisms and content from projective geometry in that the two lines 
e.g.~may be interpreted in terms of generators of a hyperboloid or 
general ruled surfaces or -- associating e.g.~focal properties -- in
terms of ray systems \cite{kummer:1866}. For now, we leave those details
for upcoming discussions because the current focus is to supersede or
unify the technical frameworks used so far.}.

\subsection{Projective Geometry of $\mathbb{R}^{3}$, Points and Coordinate Systems}
If we identify the four points $(x_{1}$, $x_{2}$, $y_{1}$, $y_{2})$ of 
our picture individually, because having skew lines, no more than three
points lie in each plane. Hence we are already very close to delivering
a coordinate definition in projective 3-space. The only additional 
information necessary to fix the coordinate system is an unit point $E$,
and -- having introduced such a point $E$ -- we can proceed with well-known
coordinate geometry, algebra and analysis \cite{klein:1928}. The tetrahedron
given in figure 1d can be identified as the fundamental coordinate tetrahedron,
and we may apply classical projective geometry with real transformations 
(or 4$\times$4 real matrix reps, however, we have to take care about the
transformation rep, e.g.,~its rank and further rep properties).\\
In order to lead over to the next subsection, we introduce an unit point
$E$ in a manner that the coordinates of the four points $(x_{1}$, $x_{2}$,
$y_{1}$, $y_{2})$ in figure \ref{fig:fig1} are mapped to the coordinates
((1,0,0,0), (0,1,0,0), (0,0,1,0), (0,0,0,1)). However, it is important to
mention that alternatively instead of starting with the four points like
before one can also identify the six {\it sides} of the tetrahedron and 
use them as line coordinates in a so-called six-vector rep of a line\footnote{
Here once more, we could follow much deeper concepts from projective geometry
e.g.~by interpreting each of the two skew lines as an axis of a sheaf of 
planes thus intersecting the other line in a point series, i.e. we obtain 
this line being represented as point set with appropriate projective 
relations. Two conjugated lines (related by reciprocal polarity) lead to
ray systems of $1^{\mathrm{st}}$ order and class \cite{kummer:1866}, 
\cite{dahm:MRST3} and to null systems. However, that's beyond scope here.}.

\subsection{SU(4) and SU$*$(4)}
\begin{figure}[h]
\includegraphics[scale=0.5]{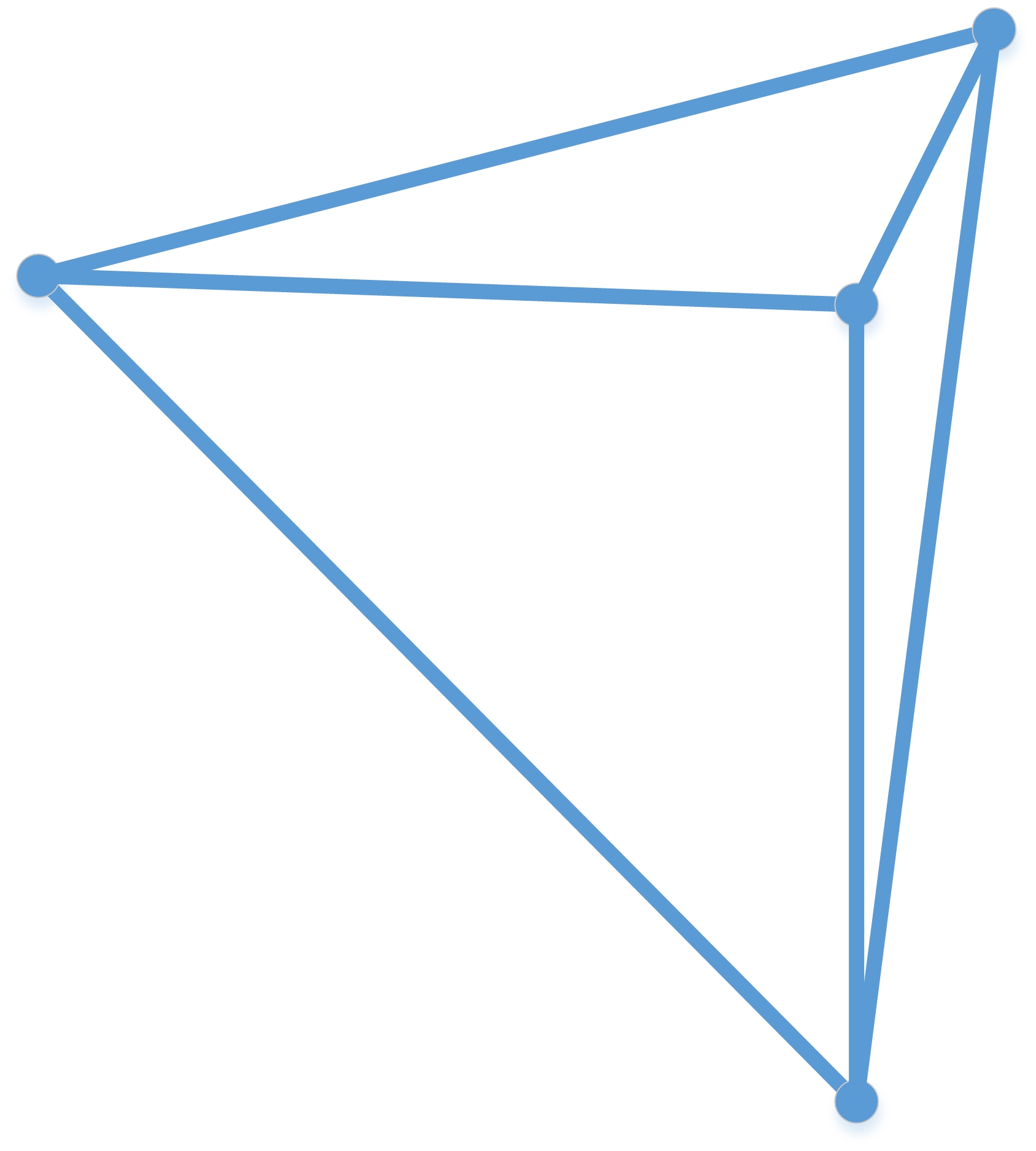}\hspace{2cm}
\includegraphics[scale=0.5]{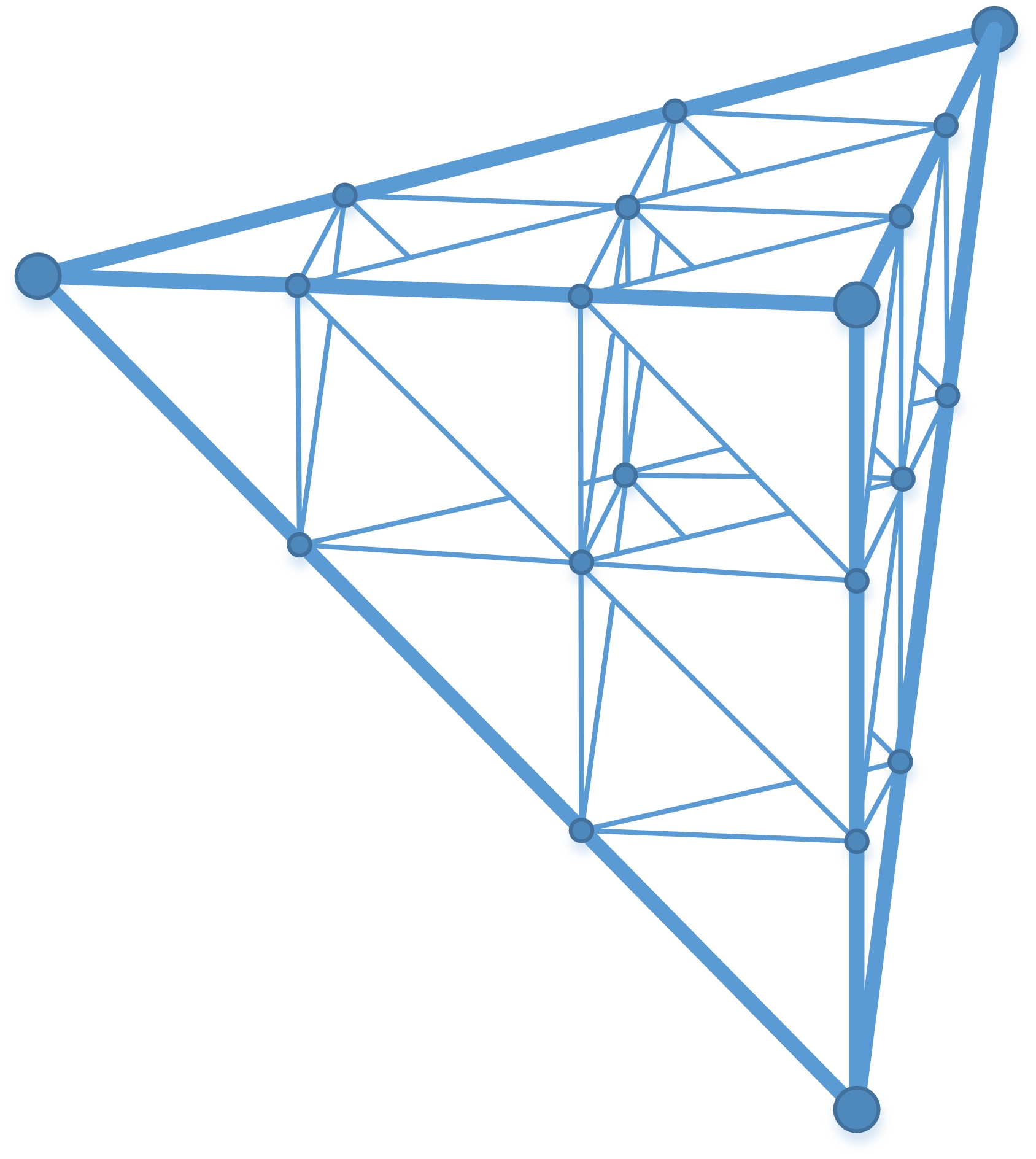}
\caption{SU(4) multiplets \protect{\underline{\bf 4}} and 
\protect{\underline{\bf 20}}}
\label{fig:fig2}
\end{figure}

Now in \cite{dahm:diss}, appendix F.6 (see also \cite{dahm:1995}), 
we have constructed various representations of SU(4) which we may use
immediately based on the above given points of the tetrahedron. Because
SU(4) ($A_{3}$) is compact and of rank 3, we can transform the roots to
a real 3-dim rep and draw pictures of the reps (see \cite{dahm:diss}).
The fundamental rep \underline{\bf 4} of SU(4) yields a tetrahedron,
so we feel free to identify the root space diagram \underline{\bf 4}
with the fundamental tetrahedron of real projective 3-dim geometry
{\it and vice versa}! Moreover, due to Young diagrammatics we may
understand various representations as being build out of compositions
of tetrahedra, a special case is the representation \underline{\bf 20}
comprising four nucleon and sixteen delta degrees of freedom which 
we have discussed at various places before (see references). We have
stated a surprising 'selfsimilarity' of the representations (see 
figure \ref{fig:fig2}), and, sloppy speaking, \underline{\bf 20} is 
a 'cubic' of \underline{\bf 4} due to the third-order symmetric tensor 
product\footnote{So with respect to $3^{\mathrm{rd}}$-order curves, 
it is natural to expect (up to a common/overall normalization factor
of the states) eigenvalues of $\pm 1$ and $\pm 3$, see \cite{dahm:diss}
and \cite{dahm:MRST3}.}.

Acting with (general) complex transformations, of course, transforms
the individual points of the tetrahedron \underline{\bf 4} to general
complex (4-dim) coordinates, however, we know that \underline{\bf 4} 
and \underline{\bf 20} both are irreps of su(4), so for very special 
choices of complex transformations (i.e. su(4) transformations\footnote{Care
has to be taken when acting with the {\it group} SU(4) because this
involves anticommutators as well which complicates the situation 
considerably. Especially, also ternary products like Lie and Jordan
triple systems (based on commutators and anticommutators) enter the
scenario and yield additional information and constraints (see 
e.g.~\cite{dahm:2008} and references), although the consequences 
are not independent from the underlying algebra or group.}) we transform 
tetrahedra into themselves\footnote{A detailed discussion of the topics
mentioned here has been given in \protect{\cite{dahm:QTS7}} but due to 
some time problems this is not published yet. Therefore, a short summary 
is given in appendix \ref{app:SU4reps}.}. So obviously, here we can relate
SU(4) Lie theory with (real) 3-dim projective geometry. To relate to 
non-compact SU$*$(4), we have to keep in mind that SU$*$(4) represents
the complex embedding of quaternionic SL(2,$\mathbb{H}$), and that we
began by embedding two independent quaternions ('spin' and 'isospin' 
Lie algebras), i.e. two $A_{1}$ into $A_{3}$. Now with respect to such 
scenarios e.g.~in elliptic geometry we have cited already some relevant
literature (see references in \cite{dahm:MRST3}), here we want to 
emphasize only Study's work and Lie's line-sphere transformation as
examples. The easiest interpretation at this time is to understand 
the so-called 'six-vector' of the line in terms of two 3-dim 'vectors'
(its 'polar' and its 'axial' part) and require conservation of the 
3-dim vector squares, even in the complex case\footnote{This allows 
to represent and investigate typical properties of 3-dim 'vector' 
reps, of course, within its associated (euclidean) interpretation,
'vector' pictures like 'parity' and 'chirality' and appropriate 
counterparts when transferred to complex rep spaces e.g.~in the 
context of \protect{SL(2,$\mathbb{C}$)}. However, here we do not 
want to discuss such features or detailed reps of 'chiral symmetry'.}.
SU$*$(4) on complex rep spaces or SL(2,$\mathbb{H}$) as general real
transformations of two real quaternions (and isomorphic to the Dirac
algebra) thus can be used to represent certain geometrical transformations
and behaviours, especially when mapped/transferred to line or projective
geometry.

\subsection{Projective Geometry from Scratch}
It is even possible to look at the tetrahedron from a strictly synthetic
viewpoint in terms of line geometry \cite{vonStaudt:1856}. This can be 
easily seen by extending the basic planes of the tetrahedron in all 
spatial directions. So a general line\footnote{The line shouldn't lie 
in one of the planes and shouldn't hit one of the vertices of the tetrahedron.}
in $\mathbb{R}^3$ will hit the planes in four different points. Now
von Staudt showed \cite{vonStaudt:1856} that the anharmonic ratio\footnote{German: 
Wurf, Doppelverh\"{a}ltnis} of the four intersection points of this 
line with the tetrahedral planes and that of the four planes each 
comprising this line and one of the four tetrahedral points is the 
same as long as the order of elements in the anharmonic ratio is the
same (\cite{vonStaudt:1856}, Erstes Heft, \S 2, especially numbers 35
and 36). So the 'playground' is set for transformations respecting
anharmonic ratios or investigations of the set of the $\infty^{1}$
anharmonic ratios, i.e.~we are in the center of projective geometry.
The more general theory of quadratic Complexe was developed in 
\cite{plueckerNG:1868} or in Reye's 'reprise \cite{reye:1866} (there
especially part 2) of von Staudt's work.

Historically, and that's the basis to propose line geometry to describe
dynamics as well, Complexe (and especially tetrahedral Complexe) appeared
in various contexts. When considering the movement of rigid bodies
around one of its points, the (infinitesimal) rotation axes build a 
(second-order) cone. Further examples are their Dreibeins of (orthogonal)
axes of inertia which are attached to each point of the body and as 
well attached to each point of space when 'moving space' with the body, 
and also the related normals to confocal surfaces of second grade which 
were parametrized by
\begin{equation}
\label{eq:confocal}
\frac{x^{2}}{a^{2}+\lambda}
+\frac{y^{2}}{b^{2}+\lambda}
+\frac{z^{2}}{c^{2}+\lambda}
=1\,,
\end{equation}
$\lambda\in\mathbb{R}$ representing the 1-dim parameter of the set
of surfaces. And we may use quadratic Complexe as well to approach 
'light cone' or sphere reps in point coordinates (see \cite{dahm:MRST3} 
for a reference to \PL and work of Binet and Dupin). In plane coordinates 
$u$, $v$, $w$, eq.~(\ref{eq:confocal}) reads as
\[
(a^{2}+\lambda)u^{2}
+(b^{2}+\lambda)v^{2}
+(c^{2}+\lambda)w^{2}
=1\,.
\]
Being linear in $\lambda$, we can attach a normal in the contact point
of the surface with the plane, and we thus arrange/obtain a unique 
mapping of planes in space to normals. Similar investigations have 
been done for infinitesimal transformations, for the Complex of tangents
related to (infinitesimal) motions and related second-grade cone(s). 
More general, we can map two (rigid) bodies pointwise onto each other, 
or extend this mapping to projectively mapping points in space \cite{reye:1866}.
If we use lines to connect the original points with their individual 
images, the lines constitute a tetrahedral Complex, so our picture 
in figure 1d above is obviously a subset of the general construction
scheme. Moreover, considering normals of concentric second-order 
surfaces, which obey a constant anharmonic ratio with certain planes
including the plane at infinity, leads to a tetrahedral Complex. This
holds also for the normals of confocal surfaces of second order (see
e.g.~\cite{vonStaudt:1856}, \cite{reye:1866} and references therein).
We have mentioned already the well-known feature of e.g.~Lorentz 
transformations to leave (affine euclidean) normal coordinates $x$, 
$y$ with respect to a $z$-axis of motion invariant \cite{dahm:MRST3}.
This can be translated back to transformations of homogeneous or 
projective coordinates, normal planes (or null systems), or to the 
line coordinates themselves. So asking for invariance of certain 
ratios of homogeneous line coordinates or invariance of some line
coordinates themselves leads to subsets of projective transformations
and may be considered separately.

Having based our discussion so far on quadratic Complexe (i.e.~Complexe
of grade 2) and especially tangential and tetrahedral Complexe, on one 
hand, we feel very comfortable to apply the full framework of projective
geometry, including dynamical treatment of points, second-order curves 
and surfaces, polar relationships, etc.~in a more general and purely
geometrical framework. On the other hand, this approach provides and
satisfies the treatment of the requirements of \cite{ehlers:1972} in
a more general description than only using points and point manifolds.
Of course, we are aware of transfer principles mapping objects like 
lines, Complexe or spheres to points in higher-dimensional spaces\footnote{
E.g.,~the \PLL-Klein quadric, Laguerre geometry, cyclography, etc.},
however, here we want to take up at first the 'classical' position
of a real 3-dim projective geometry and mention the enormous wealth
of this description (without switching to more complicated approaches
which usually introduce additional ad hoc-assumptions or axioms on
manifolds). So using \cite{ehlers:1972} for the moment as a guideline
to state the necessary (minimal) theoretical requirements of relativity,
line and Complex geometry provide a unification basis for us with special 
emphasis on Complexe of first and second grade \cite{plueckerNG:1868}, 
\cite{reye:1866}.

\begin{figure}[ht]
\includegraphics[scale=0.5]{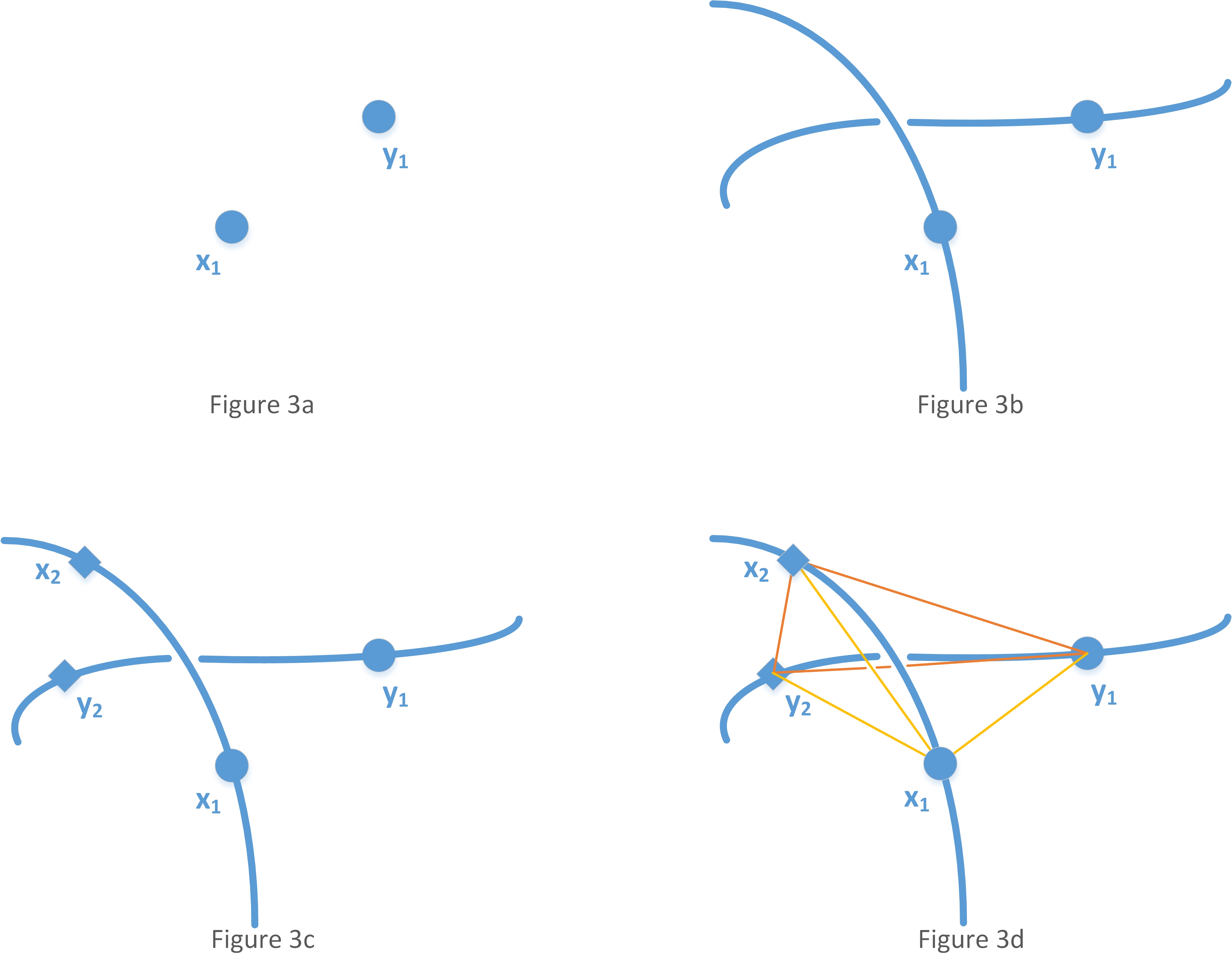}
\caption{Higher-order motion}
\label{fig:fig3}
\end{figure}

Before closing this section, it should be mentioned that we may replace
the straight lines in figure \ref{fig:fig1} by curves as well, see figure
\ref{fig:fig3}. If the curves themselves respect further properties or 
obey further constraints, e.g.~with respect to polar relations, null 
systems, conics with projective generation, Complex curves, intersections
of higher-order surfaces, higher-order curves, etc., this enriches the 
given picture considerably but the treatment is completely possible 
within the approach presented here.

\section{Outlook}
After having presented various representations of projective 
geometry, our roadmap is fixed towards looking deeper into
dynamics from the viewpoint of projective geometry with special
emphasis on line and Complex geometry. We have mentioned few 
aspects already in the last section and in \cite{dahm:MRST3}.
At the time of writing, we are convinced of having described
and represented various aspects of projective geometry so far
in terms of different group models and/or representations, so
it is worth to look deeper into dynamics formulated in terms 
of line and Complex geometry.
Last not least, we want to thank Bernd Schmeikal for deep and
enlightening discussions at the Goslar conference 2015 and
in Vienna in summer 2015, for his hospitality, interest and
time there during my visit, and especially for his great work 
on logic based on fourfold base elements \cite{schmeikal:2015}
which we want to see related to the two orientable lines or 
line elements in figure 1 (or figure 3). Moreover, it is a 
great pleasure to thank George Pogosyan and his team for the 
great conference athmosphere and organization in Yerevan and
for the enormous hospitality while staying there.

\appendix

\section{Short Remark on SU(4) Representations}
\label{app:SU4reps}
We want to summarize briefly some SU(4) rep facts as far as
they are -- to our opinion -- related to our reasoning with 
respect to projective geometry\footnote{Conference talk at
QTS 7, Prague, 2011, publication upcoming soon}.

The rank-3 group SU(4) (or $A_{3}$) has three fundamental 
representations \begin{Young}\cr\end{Young} or $[1,0,0]$,
\begin{Young}\cr\cr\end{Young} or $[1,1,0]$, and 
\begin{Young}\cr\cr\cr\end{Young} or $[1,1,1]$, of dimensions
4, 6 and 4, respectively,
(see e.g.~\cite{dahm:diss}, appendix F.6, or \cite{lichtenberg:1970},
p. 99/100) and we may perform standard transformation operations 
with the reps above or construct interactions or construct 
invariants etc. Now from Young diagrammatics, we know that the
reps are antisymmetric with respect to vertical rows (due to
the permutation group construction scheme). Moreover, we know
that $[1,1,0]$ is selfconjugate and of dim 6 whereas $[1,0,0]$ 
and $[1,1,1]$ (each of dim 4) are mapped to each other by 
conjugation, and we thus work with an involution.

Now identifying \underline{\bf 4} $\sim [1,0,0]$ with a point rep,
the antisymmetric 'products' of point reps in case of $\sim [1,1,0]$
are equivalent to the very definition of a line rep (see \cite{dahm:MRST3}
for the analytic expression) whereas $\sim [1,1,1]$ maps to a 
third-order (sub-)determinant which complies to the 'standard 
definition' of plane coordinates $u_{\alpha}$, $0\leq\alpha\leq 3$,
in terms of (projective) 4-dim point coordinates of $\mathbb{R}^3$,
i.e.~we find natural associations to \underline{\bf 6} and the 
second \underline{\bf 4}. Whereas the higher-order products can
by calculated/represented by Young diagrammatics, here as a 
second approach towards the background rep theory, we want 
to mention a special, but established aspect of projective 
geometry (see \cite{doehlemann:1905}, I \S 3, Nr. 8), valid 
for general polyhedrons with respect to duality in 3-dim space, 
here however applied with respect to the tetrahedron. There, 
denoting by $e$ vertices, $f$ areas and $k$ edges, the dual 
(or 'reciprocal') polyhedron consist of $e$ areas, $f$ vertices
and $k$ edges. Now, in the case of the tetrahedron, we thus 
find/recover the mappings $4\longleftrightarrow 4$ and 
$6\circlearrowleft$, which reflect in the three fundamental
reps \underline{\bf 4}, \underline{\bf 4} and \underline{\bf 6},
and emphasize their identification with point-, area- and 
line-reps (edges).

So we feel authorized to use projective (3-dim) geometry from scratch,
and subsume the group theoretical approaches and reps to cover certain
analytical facets thereof. As with respect to the usual discussion of
complex numbers we point to the LONG discussion in (geometric) literature
and with respect to geometric interpretations. And no, we do not want 
to discuss and understand complex numbers only from the (contaminated)
viewpoint of complex analyticity and differential geometry.

Last not least, we want to point to the use of the background discussed
above in the case of self-polar tetrahedra in coordinate systems and 
especially in line and Complex geometry (see e.g.~\cite{klein:1928})
as well as the well-known context of \cite{kummer:1866} to focal surfaces
which we can relate immediately to vertices in QFT. The very definition 
of a focal surface (\cite{kummer:1866}, p. 5) yields the definition of 
a standard vertex (1 line or momentum $\longleftrightarrow$ 2 lines or 
momenta), so for $n=2$ and $k=2$ (i.e.~ray systems of order $2$ and 
class $2$) we may apply this framework in that we identify a QFT vertex 
as being a point of the focal surface and proceed with line/Complex 
geometry instead of (sometimes) mysterious 'quantum' argumentation in
terms of point manifolds\footnote{Reading \cite{kummer:1866}, it is noteworthy
to point to theorems 33 (\S 7) and 38 (\S 8) for later use as well as
to emphasize the common ground with Jacobi's and Hamilton's classical
approaches to dynamics.}.

\end{document}